# Datalism and Data Monopolies in the Era of A.I.: A Research Agenda

Catherine, E.A., Mulligan*

Phil Godsiff

The increasing use of data in various parts of the economic and social systems is creating a new form of monopoly – data monopolies. We illustrate that the companies using these strategies – Datalists – are challenging the existing definitions used within Monopoly Capital Theory (MCT). Datalists are pursuing a different type of monopoly control than traditional multinational corporations. They are pursuing monopolistic control over data to feed their productive processes, increasingly controlled by algorithms and Artificial Intelligence (AI). These productive processes use information about humans and the creative outputs of humans as the inputs but do not classify those humans as employees, so they are not paid or credited for their labour. This paper provides an overview of this evolution and its impact on monopoly theory. It concludes with an outline for a research agenda for economics in this space.

## 1 INTRODUCTION

Digital technologies now play a fundamental role in our global economy; few areas of society remain unaffected. In many instances, it is not easy to participate in economic life without using digital tools – both as an employee and a citizen (Foroohar, 2020). Previously, Multi-National Corporations (MNC) have applied ICT to improve productivity of pre-existing business processes and systems (OECD, 2018) – so-called "Big Tech" firms, however, instead use data as a fundamental input to their productive processes; vast infrastructures have been established to track, trace and store significant quantities of data to the point that companies such as Meta (formerly Facebook), Amazon, Google, Microsoft and Apple have received their own moniker – MAGMA (previously FAGMA). In addition, newer companies such as Open AI are leading a new hunt for data sources to train their large language learning models (LLMs). Big Tech is therefore pursuing a different type of monopoly control than traditional MNCs; they are pursuing monopolistic control over data to feed their productive processes, which are increasingly controlled by Algorithms and Artificial Intelligence (AI) rather than human employees. This paper illustrates that the existing theory around monopolies is incorrectly focused to handle the emerging data monopolies created by Data Monopolists – or *Datalists*, rather than "Big Tech". We propose brief research agenda in this important area.

Significant amounts of focus have been placed on the privacy invasion associated with the collection of data about individuals by MAGMA, e.g. (Wang et al., 2011). However, a more extensive set of economic issues also warrant investigation: the value of goods and services is no longer necessarily created solely *by* human labour but is now also created by algorithms analysing data *about* humans, human activity (e.g., creative outputs on blog posts of video and image sites), and the use of different Datalist services (Birch et al., 2021). Data in capitalist modes of production is, first and

Imperial College Business School, c.mulligan@imperial.ac.uk

foremost, a means to capital's end, and algorithms are the machines of the digital age; humans are now embedded into the machines through the use of data about them as users - "You are either in the data, or you are working on it" (Microsoft employee, quoted in Roose, 2021).

The increasing control exercised by Big Tech over the activity of other economic actors and their increasing size has brought them under intense scrutiny by regulators and public authorities: "Market tipping" (Furman, 2019), positive network effects, economies of scale, strategies of external growth, including "killer acquisitions" of current or nascent competitors (USSC, 2020) have all been put forward as reasons for their exponential growth. Many have proposed that Big Tech companies display monopolistic behaviours (Petit, 2020) and even suggested that regulation is implemented (Lande, 2021). Traditional measures of monopoly power are usually based on price, e.g., the price elasticity of demand (Sanchez-Cartas, 2020), price discrimination ((Sanchez-Cartas, 2020) or earning higher than regular profits (Sanchez-Cartas, 2020). However, these measures have proven difficult to apply to Big Tech companies (OECD, 2018), not just because they are often working with data but because their global reach means assessing a company's ability to control price is more challenging.

US companies alone are sitting on approximately $2 Tn in cash (Economist, 2016); rather than continue to pursue control over markets and profits as traditionally understood by monopoly capital theory (MCT). Therefore, Datalists have instead sought monopsonist / oligopsonistic control over *resources* required in the digital economy, namely data and algorithms. Datalists – in contrast to traditional large MNCs - have focused on amassing large amounts of data about individuals, machines, and companies and turning them into productive resources they can monetise. We need to differentiate between large companies pursuing traditional monopoly strategies and Datalist companies pursuing a data monopoly strategy –this is directly relevant to MCT.

There have always been large companies - Foster and McChesney (2012) demonstrated that during the Global Financial Crisis (GFC), the top 200 corporations in the US accounted for about 30 per cent of gross profits in the economy - up from 13% in 1950. Global monopoly firms also bounced back faster than other sectors of the economy after the GFC – "there is little doubt that this recovery of the giant firms was related to their monopoly power, which allowed them to shift the costs of the crisis onto the unemployed, workers, and smaller firms (Foster et al., 2011). The IPS (2000) also identified that - after comparing corporate revenues with the gross domestic product (GDP) - 51 of the largest 'economies' in the world were corporations versus 49 national economies. Grullen et al. (2019) illustrated that over 75% of US industries had experienced an increase in concentration levels. Over the past two decades, most US firms are making more money than they used to, and more firms have become very profitable – indeed, Return on Capital is at near-record levels (Economist, 2016). ICT has been crucial in this monopoly capital accumulation over the last 20 years (Nolan, 2008). ICT has formed a foundational part of the reach of all MNCs – many Fortune 500 companies have large-scale, in-house technical teams ensuring 24x7 global operation between geographically dispersed units (Nolan, 2009). MCT, therefore, works exceptionally well if we are talking about MNCs that are pursuing monopoly by traditional means – namely, the sale of physical goods and services.

Big Tech companies, however, are enjoying not just significant growth in their profits *but growth that far outstrips even other substantial multi-national companies,* as illustrated in Fig 1 and Fig 2. We illustrate that *beyond* the typical behaviour associated with monopoly companies within MCT; MAGMA and others like OpenAI have also pursued control over crucial inputs in the modern economy – namely, data associated with end-users that they convert into an input to their production processes. These data, in turn, power technical intermediaries such as supply-side platforms, demand-side platforms, and advertising exchanges that are essential components of personalised, highly tailored advertising towards individual end-users. Google and Meta, for example, are now almost entirely vertically integrated along the whole chain of digital advertising (ADLC, 2018); Datalist companies have succeeded where others have failed – in the application of data. What started with advertising has now moved far beyond advertising, as illustrated in the following sections.

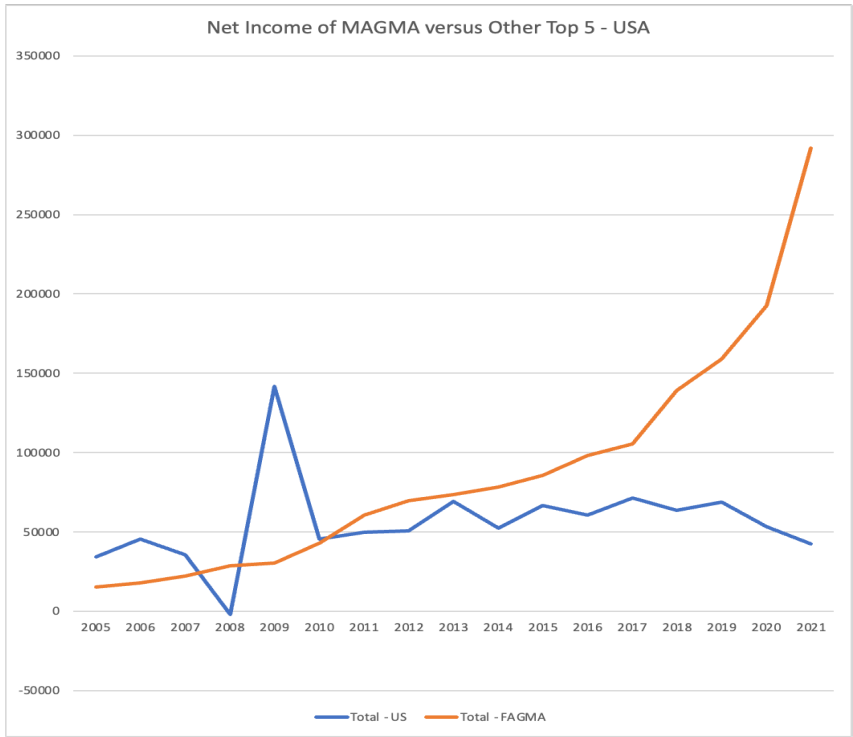

Figure One: Net Income of MAGMA versus other top five companies

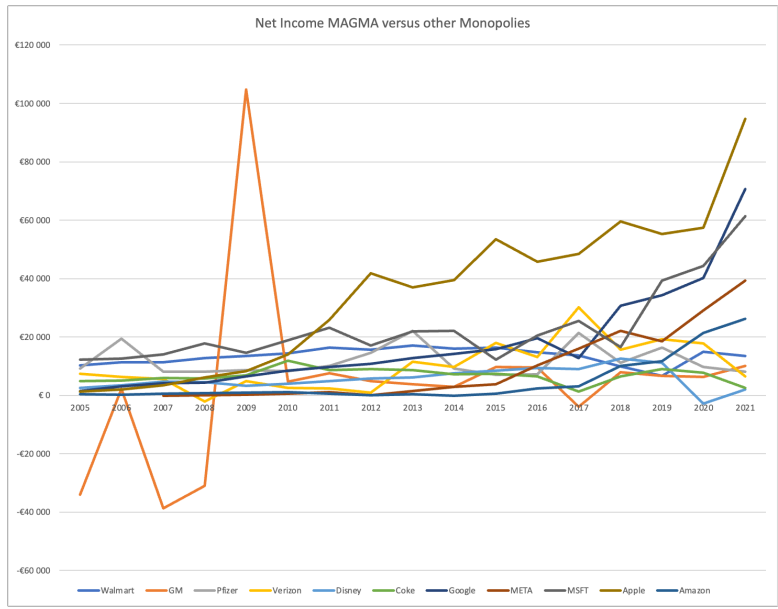

Figure Two: Net Income MAGMA versus other monopolies

## 2 BIG TECH AS DATA MONOPOLISTS - DATALISTS

Within this paper, we focus on one aspect of Big Tech strategies – namely, their monopoly over data; we define "Big Tech" as the interaction between two things within the digital economy: digital data and algorithms. Digital data and algorithms are the basis of Datalists' growth. We provide a unique name for the company pursuing this type of monopoly over data – Datalist - to differentiate them from traditional monopolists, who often seek market dominance or price control instead. As we will discuss, they are pursuing different types of monopolies, and we need to develop new means to measure monopolistic activity in the Datalist era.

**Digital or Liquified Data**: "Data" is not new; the earliest known data storage was approximately 10,000 BC. There is a need to differentiate, however, between what we are calling analogue data (siloed, stored on separate systems or on paper, proprietary and complex if not impossible to share easily) and *digital* data – this is data that can be exchanged between machines with little or no human intervention. Previously, a company's analogue data about its customers and other information were kept within the boundaries of the firm either in computer systems that were not open to others or in filing cabinets. Generally, this data was the product and result of other physical business processes – processes that the company would perform whether the data was collected. In approximately 2004, however, a shift occurred in ICT - data became more fully "liquified" (Normann, 2001) and began flowing *across* firm boundaries via technical interfaces called Open APIs (Mulligan, 2012). Open APIs are an extension of Application Programming Interfaces (APIs). Programmers initially used APIs to connect different parts of a technical system internally to the company they were working in or within a closed set of interfaces that meant the data exchanged via such interfaces remained inside the boundary of the firm. The firm retained the data it had created about its customers, products, etc.

Open APIs, meanwhile, were used to expose data outside of the boundaries of a firm (Mulligan, 2012); data that flows over Open APIs is used *between* companies or between a company and a user. Some of the most famous Open APIs include Twitter's 'Firehose' and Facebook's Developer APIs. Open APIs turned these platforms from merely communication mechanisms into something fundamentally different - they became producers and aggregators of valuable data sets in the newly emerging data economy.

Moreover, it turned end-users into *inputs* into a production process they do not earn a wage from or share in the profits from. The end-users became production mechanisms for large-scale data supply chains (Spanaki et al., 2018). In some instances, Big Tech even require end-users to purchase devices so that data can be mined from the users – e.g., Google's Android handsets or Amazon's Kindle. Every time a user searches on a Google device or searches for and buys a book on a Kindle, the data about how a user is consuming not just the product they are directly engaging with but all products they have engaged with is collated, aggregated, and analysed. Big Tech companies are therefore expropriating data and using it as part of the production process from people who have never been - and never will be – "employees". This changes the fundamental nature of work and redefines the very foundations upon which economic relationships have been built and understood in economic theory. This is becoming increasingly clear in the era of AI, as the full impacts of the Datalist model of our economy become increasingly clear with strikes by the Screen Actors Guild – American Federation of Television and Radio Artists (SAG-AFTA), and the Writers Guild of America (WGA) highlighting the increasing role of *data* in media production. The most recent AI proposal suggests scanning background performers, paying them for a day's work, and granting their companies perpetual ownership of the scan, image, and likeness for unrestricted use without consent or compensation (Frenkel, 2023). The use of data sources found on the internet – stories, news articles, artworks, message board posts, photographs, and other creative content created by individuals and shared on the internet has been scraped to feed to the LLMs (Frenkel, 2023). Scraping is taking data from the public internet by cutting and pasting the contents of websites, blogs, message boards and social media sites. This scraping is mostly done without consent, citation, or payment. It is not just creative producers who are being reduced to mere inputs to the productive process of the LLMs; coders are suing Open AI after their code was used to train OpenAI's programming assistant, which helps other coders to develop software.

These interfaces enabled third parties unconnected with the Big Tech company to access data housed in the platforms and build applications on top. The world's first Open API of this nature was launched with Facebook in 2004. These Open APIs enabled the transformation of data about end-users into a productive resource. A key aspect is that most of the companies in question – MAGMA in particular – launched free services (Android et al. Stores, Amazon Marketplace) solely to gather data about users of their services. We place this as a demarcation of a new type of economic interaction emerging around data; and the rise of Datalists - companies like MAGMA and others, such as OpenAI, whose sole aim is to monetise data collected from various sources. The significant number of products and services launched by these companies has been solely about collating – and monopolising – data rather than solely about selling the product or service itself. This is a crucial distinction between traditional monopolies and datalists.

**Algorithms**: Algorithms are pieces of code written to take data, process it and produce outputs for decision-making. A constant data supply is critical to creating market-leading algorithms in large-scale machine learning and Artificial Intelligence systems. Algorithms – the new machines of the digital economy – work together with data. Without algorithms and data, the large data centres and other types of physical machinery would be of little actual use; the physical infrastructure exists solely to collect and control data over large sets of users, remote devices, and websites visited and to extract patterns (ADLC, 2018, ADLC, 2019b). Individual data points cannot extract patterns or relationships between variables (ADLC, 2016).

Datalists have therefore succeeded in capturing significantly higher than average profits due to captive audiences on their free services and the algorithmic capabilities to turn data about a user's *consumption* of their services into not just an asset – but a productive resource from which they can generate revenues. They are therefore creating a *monopoly over data about users and from users* rather than focusing on the production processes of physical goods.

Big Tech built on data monopolies are not purely a Western phenomenon. Within this paper, we focus on MAGMA rather than, for example, the Chinese competitors – Baidu, Ali-Baba and Tencent (BAT) because MAGMA are more global, less geographically and politically confined, and therefore exhibit one key critical aspect of data monopolies – the continued pursuit of broader and more sources for data (World Bank, 2010) globally.

Often monopoly power is investigated from the perspective of industry concentration. However, global MNCs are now also multi-industry in nature, such as Amazon, which has used its technology to significant effect, sequentially transforming from a bookstore to a general department store, electronics store and even a grocery store – all driven through its deep insight into aggregate customer demand –based on the data it was collecting and using for insights and knowledge, and supplier success; through knowing what products were successful in their own customers' catalogues – Amazon were able to use aggregates of such data to create 'own brand products that compete directly with Amazon's marketplace customers (Cremer et al., 2019).

Through data accumulation and increasingly advanced levels of digitalisation across industries, monopoly capital has the means to separate itself from the necessity of purchasing labour as an extension of capital. Monopolistic data accumulation enables the disembodiment of capital accumulation from the labour processes. Once fully understood, this notion challenges our understanding of the interactions between monopoly capital, workers, and society; a new type of capitalism is emerging (Pagano, 2014).

There is a clear delineation between companies' strategies of applying traditional ICT traditionally and their implementation of a datalisation strategy (DAS). This shift started in 2004 with the release of Facebook, the first company in the world to turn users' use of their platform into an asset (Wang et al. l, 2011, Birch et al., 2021). Other Datalist companies followed suit at later dates – but there is a clear link between when they started their Datalist journey and growth in profits.

## 3   MONOPOLY CAPITAL THEORY AND DATALISM – BRIEF LITERATURE REVIEW

This section briefly reviews the MCT literature, explores its predictions and policy recommendations, and illustrates its relevance to Datalism.

MCT has a long and distinguished heterodox history, reaching, according to some (Sawyer, 1988), as far back as Lenin and with clear antecedents in the work of Marx and his analysis of capitalism. Indeed, there is a Marxist heritage to much MCT literature and its criticism of the impact of monopolies (Sawyer, 1988), especially from those based in the US (Auerbach & Skott, 1988). This pedigree places MCT firmly within a more classical 19th-century economic school approach concerned with growth, distribution of resources and power, and naturally within the remit of political economy, rather than current dominant economic schools more concerned with markets and price.

MCT outlines that those monopolistic tendencies due to firm and market concentration lead to deleterious consequences for a wide variety of stakeholders, on which most authors in this field are broadly agreed. We may group these effects into three main areas: "economic"-primarily concerned with welfare losses, price, and productive efficiency, "reduced innovation", - leading to stagnation; and "social" - those primarily concerned with negative distributive impacts and exercising power and control.

Economic and market effects include notional welfare loss, higher prices and lower production levels based on a comparison with an ideal form of perfect competition, creating - and then wasting - surplus through managerial costs or advertising. In addition, it is seen to reduce innovation, leading to stagnation in the economy, creating economic and non-economic barriers to entry. MC also restricts access to or monopolises resources; restricting personal, corporate, and national development; seeking the potential lowest cost option through identifying global opportunities, resulting in potential de-industrialisation. Negative Social impacts, meanwhile, include a tendency for surplus to be "wasted" or diverted towards rentiers leading to increasing inequality and reductions in aggregate demand, inequitable distribution of the surplus, and abuse of market and state power (Sawyer, 1988).

An additional feature of the MCT school of thought is its emphasis on the finance industry's growth beginning in the 1980s as a medium of investment or economic surplus absorption (Foster & Magdoff, 2009; Lambert, 2011). As Robinson (1933) discussed, monopolistic behaviour in an oligopolistic context is rendered possible by the appropriation of resources necessary for commercial development: "More and more, the great firms have a foot not only in many markets but in many industries, in several continents, the connections between their various activities being neither in know-how nor in marketing but merely in financial power." Similarly, as Cowling and Tomlinson (2005) predicted, broadscale globalisation leads to the negative consequences of the ability to trade off investment between countries, decreased international competition, and quasi-state power and influence. These predicted impacts of globalisation are immediately relevant to discussions about Big Tech and data.

One of the most significant issues associated with the modern era is, as Cowling and Tomlinson (2015) predicted, the increasing transnationalism which has resulted in a more significant imbalance of power between capital and labour since it facilitates a process of divide and rule and therefore tends to hold down wage costs (Autor, 2020).

The monopolistic firm accretes power over society as well as over the market. "For instance, Cowling and Sugden (1998, p. 67) state that a modern, large corporation is the means of coordinating production from one centre of strategic decision-making': whether transactions are market or non-market is not the issue (in Cowling and Tomlinson, 2005, p42), which is seen in today's extensive lobbying of governments and even NGOs such as the United Nations by Big Tech firms. In many instances, therefore, digital has come to act as a form of corporation coordination mechanism, allowing companies to 'hollow out' operations and instead form a web of control over the activities of others via digital means (Nolan, 2008). They are controlled by "elite corporate hierarchies from one centre of strategic decision-making: indeed, this is the essence of the modern corporation, where activities are often extended beyond the firm's legal boundaries" (Cowling & Sugden, 1998)

The economic unit was no longer a small firm "producing a negligible fraction of a homogenous output for an anonymous market but a large-scale enterprise producing a significant share of the output of an industry, or even several industries and able to control its prices, the volumes of its production, and the types and amounts of its investments" (Baran and Sweezy, 1966) - the economic unit of genuine interest was the burgeoning set of global players who would control prices, volumes and types and amounts of investments at a global scale - escaping the boundaries of national anti-trust regulation and nowhere have these set of processes been more apparent than in Big Tech. The combined forces of globalisation and financialisation combined with dramatically improving technology meant that monopolistic accumulation/concentration could seek greater profits on a global level.

Despite the original interest in MCT, by the early 1980s, the focus began to slip as economists began to prefer the perfect competition market-based view of the economy seeking productive efficiency and in renewed attempts to ensure Economics was viewed as a "Science" rather than a social science (Keppler, 1998). In his Capital and Exploitation in 1981, Economist John Weeks wrote, "The monopolies that stalk the pages of the writings of Baran and Sweezy have no existence beyond the work of those authors." Critics of monopoly capital theory claimed that the internationalisation of capital—by breaking down U.S. hegemony and making the advanced capitalist countries more vulnerable to foreign trade and capital movements—had demolished the structure of monopolistic accumulation (Weeks, 1981)

However, predictions from MCT analyses of overwhelming market power leading to undemocratic political influence on a global scale (Cowling, 1995; Cowling & Tomlinson, 2005) have been justified in the case of Big Tech. It can be posited, therefore, that MCT's fall out of favour among economists was slightly premature.

While a critical part of modern economies, technology is also missing from economic analysis (Faulkner, 2010), often treated as a 'black box' that is more or less substitutable with labour or capital. This lack of investigation is compounded in the era of digitalisation; the new types of technology that Big Tech and transnational companies are working to develop are, in fact, a new type of machine that defies being treated as a black box; it is the foundation of the economic system itself. It seems curious, therefore, that the role of technology has not taken a much larger section of economic analysis in the digital world.

Those economists looking at technology have often taken the Schumpterian route (Perez, 2010; Dosi & Grazzi, 2010). Modern mainstream economics has tended to downplay the analysis and role of technology per se in favour of approaches that reduce technology to a production function and technical progress to something that can be represented by (and that occurs between) shifts in production functions (Rosenberg, 1982; Faulkner, 2010). An illustration of this is the widespread use of the term "innovation", which aims to encapsulate all progress in science and technology but is very difficult to define and measure in the real-world (OECD, 2018). Technology is not a fungible concept but refers to specific techniques used in a specific context to solve specific problems (Franklin, 1989)

Very few economists have taken the approach of taking an external view of technology. This has been one of the most dangerous omissions regarding digital technologies - as they have a very different nature and impact than other technology. Data and Algorithms, while appearing to be a means to human ends, are ultimately nothing of the sort (Smith, 2010; Faulkner, 2010), but rather a means to capital's ends; the technology has emerged with particular technical codes built into them that do not just reproduce and legitimise existing power relations but degrades existing power relations in capital's favour even further.

We turn now to understanding how digital technology, digital data, and, more precisely, data-intensive algorithmic tools have allowed Big Tech to change the framework in which they operate. We demonstrate that the development of this new type of technology led to a significant shift around 2004 – the emergence of the Datalist.

## 4 DATALISTS – AN ILLUSTRATIVE DISCUSSION

For our discussion, we split the use of ICT within companies into two distinct sets of strategies – pre-Datalisation and post-Datalisation. Pre-Datalisation, companies pursue monopoly in traditional ways –using digital to improve existing business models and processes, data stays within the firm's boundaries. Post-Datalisation, companies focus on appropriating data outside their traditional boundaries – either from end-users' consumption of services they provide for free, their use of devices, or even from other companies or other companies' machines. Post-Datalisation, companies start pursuing a monopoly over data that can be used as productive inputs. The Datalisation era began with the launch of Facebook – the first company to effectively capture end-user data for profit-making.

### 4.1   1960 – 2004: Pre-Datalisation

Before 2004, most computational capacity was housed internally within the firm's boundaries. While the PC revolution had brought computing to end-users, this was relatively limited. PCs were not connected, and the capacity of the software on them was limited; data was effectively stuck inside the boundaries of the firm because transferring data was difficult and expensive, and the amount of data meant it was not very profitable to do so.

A significant majority of literature focuses on the role of the internet in enabling the development of new types of business models, particularly those focused on platforms (Parker et al., 2016; Gawer, 2014). Just connecting computers was not enough to change the types of monopolies - despite the development of the internet, most of the computing power was still internal to corporations. Home users of computing could not, for example, build their own banking system; to develop a company to build a bank, you still needed vast amounts of capital, land, labour, and machinery.

Before Datalisation, ICT could be viewed merely as another type of machine –different to the looms and mills of the early industrial revolution, but from a conceptual perspective, a relatively similar machine, nonetheless. It was operated by a human, who entered data into it, ran programs and performed other activities on the computer to create outputs that, more often than not, helped the company sell existing physical goods or services. ICT was an auxiliary, not a core base of the product or service itself. It turned existing business processes into 1s and 0s and enabled the established economic system to operate faster, globalise, and monopolise more quickly (Nolan, 2008).

### 4.2   The Move to Datalisation

Around 2004, however, two things dramatically transformed Digital Technologies from purely IT machines into the foundations of a new type of economy - a truly 'Digital' Economy, one based on a new type of machine and new relationships between land, labour, and capital. It was enabled by both Open APIs and computing capacity becoming freely available outside of the boundaries of the firm through an uptake of personal computers such as laptops, tablets, and smartphones that enabled the information captured about users and devices to be turned into profitable revenue streams – or datalised. The final aspect of the computational capacity revolution was cloud computing. For a relatively low price, nearly anyone could access computing power that would have previously been unimaginable for individuals or small companies. Much has been written (Parker, 2016) about how cheap access to technology has enabled a generation of new digital native start-ups; however, most have an exit strategy of selling to global digital monopolies (Pisoni, 2018).

Over time, these datalised streams between individuals, companies and even governments have increased in value as algorithms deepen the knowledge and understanding of these data sets.

Indeed, the importance of Open APIs as means to gather data that is usable as a productive resource has increased so much that in some circles, humans are referred to as "expensive API endpoints" (Rouse, 2021) acting as the middlemen between two systems; humans reduced to taking directions from one machine and inputting them into another one. Through monopolising end-user data about the consumption of their services, Datalists can have machines directing human activity rather than vice versa; perhaps no clearer example is the suggested use of AI in cinema and television (Frenkel, 2023) or in teaching; however, this is just the beginning of the Datalist era, and nearly every vocation is likely to be affected.

When viewed in conjunction with these actions, the internalisation of transaction costs via Open APIs is about more than just efficiency – it is part of the drive for monopoly control over critical assets – namely data and the algorithms that drive them. More correctly, it is a form of monopsony – but rather than being a purchaser, they have removed the necessity to purchase a critical input to their production processes through ostensibly offering services for free.

Several articles touch on the combination of Open APIs and Computational Capacity as a 'new capitalism' - often interpreted through the lens of "platforms". Platforms are indeed an essential part of the new economy – but this type of analysis does not fully develop an understanding of the new underlying relationships because it focuses on innovation (Cusumano et al., 2020) and often strategy (Gawer et al., 2014), rather than on underlying fundamentals of the process of monopolisation. The focus is on the platform to capture network effects, developers, and customers, not the conversion of data bout users into the use of data as a productive resource – data obtained from *outside* the firm boundaries through interaction with end-users.

### 4.3 The Machines of the Datalists

In all previous eras of digital technology, humans have used computers to speed up their work; for example, by increasing the speed of calculations for engineering or automating payroll runs to reduce errors. While automation may have been used to eradicate jobs, machine efforts (skills) did not develop with the human capacities using them – they displaced those jobs. In the Digital Era, however, humans are used to train the machines in question - often in the guise of games (e.g., image recognition, where people train AI machines to determine which picture is a "dog" and which is a "muffin") used to train algorithms that are ultimately designed to exceed human capability (e.g. Watson in medical diagnosis). This difference is compounded in the Digital Era with 'self-modifying' algorithms - namely algorithms designed to take data in from various sources and fine-tune themselves - in contrast to previous generations of digital technologies. Therefore, humans are no longer the means for the valorisation of the machines – instead, machines are the means for the valorisation of humans. This is illustrated in Figure 3 - a fundamentally new type of machine has emerged post-2004; data has no real value in exchange - only in use to allow machines to embody human data, capabilities, and knowledge that, in turn, allows for excluding the labour process altogether.

Today's workers now face a complex choice - they are currently training - embodying human capabilities - into the very machines that will eventually replace them (Roose, 2021); as a result, the balance of power between production inputs has been distinctly shifted in capital's favour, e.g., with "Ghost workers" now embedded in the technology industry.

### 4.4 New Machines of the Datalists and Impact on Production and Consumption

The technology industry has reached a tipping point – one where we have moved from technology delivering industrial-era business processes in a faster, more secure fashion to technology fundamentally disrupting how our economic system is formed - including how monopolies are created. Therefore, we are in the middle of the emergence of a dual economic system – one based on the traditional analogue world and traditional MC and another digital –enabled by data and computational power and datalism. We will likely have both systems interacting and co-existing alongside one another for quite some time to come. Our current economic theories used to inform policy do not yet handle the second type of genuinely digital economy.

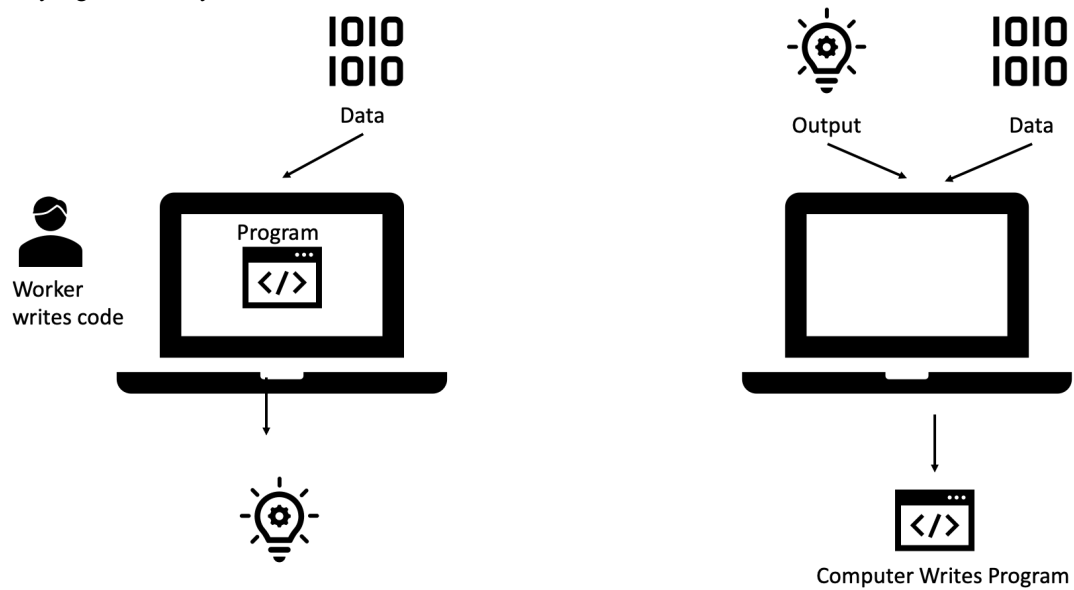

Figure 3: Pre-Datalisation and Post-Datalisation usage of computational capacity

We illustrate this concept with examples from the recent past of MAGMA in Figure 4. The stars on the graph illustrate the point in time when one of the MAGMA companies implemented their first major datalisation strategy. In 2004, Facebook launched the world's first Datalist strategy –creating and capturing data about end-users' social lives; in 2012, however, it implemented its third-party app store – enabling the same of its data products for advertising and other services. In 2007, Apple introduced its first iPhone and the associated data aspects such as its Music service, AppStore and associated data products. Google was soon to follow suit with the launch of its first Google Android device in 2008 – Google's datalisation strategy was somewhat more aggressive than either Apple or Facebook – with its ability to track and trace users while on the move, Google was able to create highly targeted advertising down to a user's location. Microsoft and Amazon were perhaps the last to join the Datalist move – Amazon implemented its most aggressive Datalist strategies since the launch of the Kindle in 2007 – it released the Amazon Echo with the AI-driven Alexa that was able to track and trace humans deeply within their home environments – not just for advertising but also for targeted product development. It also released several other Datalist products, including tablets. Microsoft meanwhile shifted to a 'data first' strategy under the leadership of its new CEO Satya Nadella in 2015; over time, Microsoft has transitioned to a data and services company.

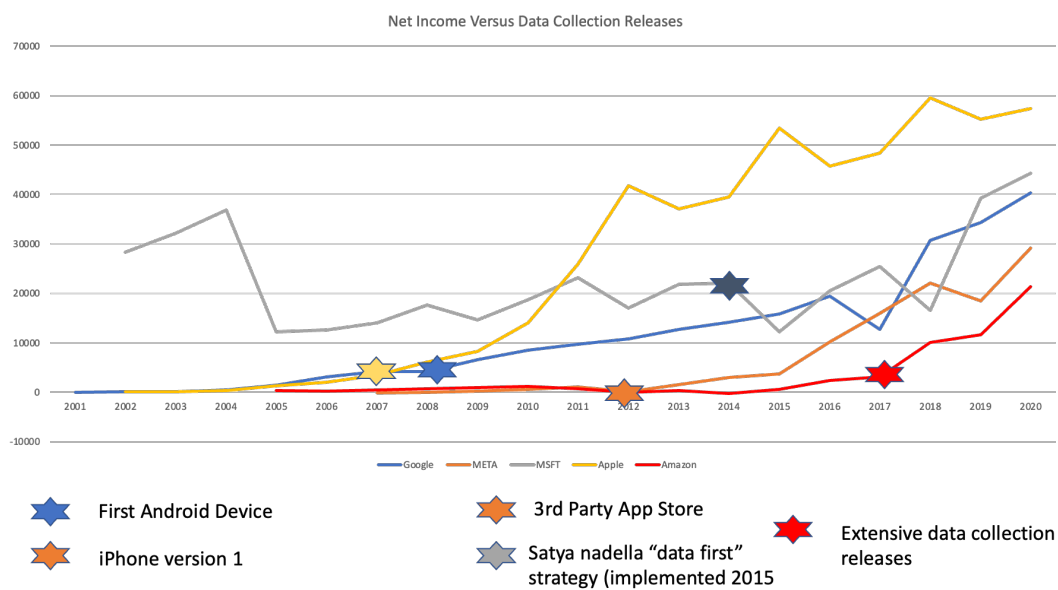

Figure Four: Uptake of data-intensive activities

Through data and algorithms, capitalists gain more even control over the work process - using humans only where necessary – and often as the interface between machines themselves. The decline of small businesses as a significant source of employment for many entry-level and less educated or less skilled workers is troubling in an age when robots, the internet or digital technology are doing an increasing amount of work once done by such a worker (Lambert, 2019). This does not, however, cover the issues of the rapidly automating back office, which is rapidly replacing even managerial staff (Rouse, 2021). A new set of control mechanisms over labour are appearing with the massive automation of white-collar or "knowledge economy" jobs; management will be replaced with data and algorithm-based management. The Datalists, far from freeing humans, have meant that they are still subordinate to the accumulation of money capital as an end in itself (Smith, 2010), but now also subordinated to data capital, feeding algorithmic machines.

## 5   A SUGGESTED RESEARCH AGENDA

While the existence and behaviour of Datalists can be consistent with the analysis and predictions contained within MCT, the role and nature of the new technologies has profound implications for both MCT and possibly economic analysis more broadly. Capitalism continues with its "complex dynamism", –exchange relations remain unequal (Dunn, 2017) and perhaps are increasingly unequal in the data economy epitomised by MAGMA. Therefore, we believe that MCT in the digital era could be usefully expanded firstly by assessing the use of data and as a critical driving force of MC's reach and scale, secondly, through focussing the unit of analysis on transnational corporations rather than national economies. Finally, the traditional approach of monopoly power being measured by a company's control over price setting must be expanded to correctly manage Datalists and their role in an oligopsonistic market for data.

While a helpful starting point for analysing Big Tech - there are some significant differences between the observations of MCT and what is happening with Datalists. Firstly, these new monopolies are generally still managed and owned and controlled by their founders or second-generation entrepreneurs with strong similarities to their founders by whom they were often selected and anointed. This reduces the impact of the argument that one of the ways to squander the monopolistic surplus is through management rather than owners siphoning it off. Control is still firmly linked to ownership. If there is a waste expenditure, it tends to veer towards the proclivities of the owners, e.g. space races and philanthropy.

Secondly, another difference is that an accepted method of disposing of (and indeed identifying) "waste" is expenditure on advertising. Bits of MAGMA, most notably META and Google, however, make money most of their income from advertising - i.e., in terms of MCT, they are benefitting from other organisations' "waste" expenditure; effectively, MAGMA monetises their user communities by enabling others to perform direct, tailored advertising to them. Therefore, it is easy to see that the Datalists' surplus arises less from productive scarcity, restriction, or resource scarcity other than perhaps from users or innovative capability. Datalists surplus instead arises more from the "collective nature of the production process" (Battistini, 2013) and the fact that they need to rely on monopolies on data from users from various sources they currently are not paying for.

Not all economists are agreed that large firm dominance restricts innovation. A privileging of the effect of monopoly and oligopoly on prices and output, along with the notion that price competition drives or should drive the economy, prevents economists from exploring that instead of competing on price, Datalists compete on access to fine-grained and nuanced data about their end-users – and compete as networks of end-users, rather than solely as individual large companies (Baumol, 2002).

Nor does everyone agree that monopoly capital leads to decreasing competition and falling rates of return. (Auerback and Skott 1988). Several argue that countervailing tendencies are provided by increased and more effective use of data and information. Carson (2002) demonstrates that under certain conditions, "monopoly output may be higher or lower than the output associated with marginal cost pricing, and higher or lower than this sector would supply in a reference economy operating under all round perfect competition.

This 'accumulation by dispossession' of data and 'commodification of labor power' (Lodhi, 2017) extends perfectly into the data economy as more and more data, information and knowledge are accumulated often for free by Datalists with services (e.g. google maps) offered both as a service but also as a means of accumulating more data from users, which they cannot use themselves as they lack access to the necessary digital and analogue infrastructure. Moreover, with the increased use of data scraped from the public internet, the commodification of labor power is increased.

An insight into what analysis of a Datalist economy might require is provided by distinguishing production and consumption systems (Subramaniam et al., 2019). Production processes become more collective. (Battistini, 2013) and the same is true of consumption processes when data about end-users is monetised through use as a productive resource

Obtaining Use information about productive systems is an essential source of competitive advantage (Subramaniam, 2019), stressing the importance of controlling "the flow of product in use information across several connected ecosystems". Moreover, this is precisely what Datalists do. However, the critical "product in use" to a Datalist is not any particular physical product that they sell to an end-user (e.g., phone, platform, software), but rather the individual *using* that product or *consuming* that service or publishing ideas and content on the service or internet. While controlling the flow of this information is essential, what is more, crucial to Datalists is being able to recombine this information in myriad ways to be able to predict, e.g., purchasing or behavioural intention to be able to create a unique advertising target or a new product/service that serves to ensure their network of end-users remain large and competitive – i.e., productive in terms of practical, usable, aggregable data.

The importance of knowledge is taken further by Rikap and Lindvall (2020), who argue that organisations like Datalists have become controllers of "corporate innovation systems", monopolising knowledge obtained internally and externally, and tightly controlling access to data. They describe MAGMA as "data-driven intellectual monopolies", leading others to describe them as constituting a new form of "intellectual monopoly capitalism" (Rikap, 2022).

The presence of intangible assets, digital data and accompanying software / creates opportunities for big tech to achieve "expansion that is both breathtakingly rapid and effectively costless" (Jacobides and Lianos, 2021 p1131). The power of intangible assets and data is recognised as a means of control by Datalists over resources rather than control over prices or output in the market (Coveri, 2021). This all has implications for economic theory. There is a growing consensus that the dominant neo-liberal economic paradigm may no longer be sufficient or relevant (Cowling & Tomlinson, 2011). The basis for a new digital economy will be data in use, not product in exchange, and will privilege the Datalists.

There have been suggestions that the technological change that brings positive network effects and blurring of the boundaries of the firm in ecosystems can reduce the relevance of accepted models and analysis (Petit, 2020). For example, how competition and firm survival operates in a dynamic digital economy with strong disruptive forces at work is different from equilibrium analysis (Petit & Teece, 2021)

Pitelis (2016) challenges the assumption of the absence of increasing returns to scale, an assumption which certainly does not hold in the data economy, with its positive network effects and where ecosystems create value that extends beyond the limits of one organisation (Pitelis &Teece, 2016)

This increased understanding of knowledge's critical role in the (digital) economy has led some to suggest that it should be considered for inclusion in the standard production function in this digital age. For example, scarcity of the resource of "genius" or skill (smith 2015 #50) which, along with digital versions of labour and capital, forms a new production function, limits the rate of growth, and can also explain low rates of recorded growth and decrease wages and interest rates (Benzell & Brynjolfsson, 2019). In addition, economics is very production and exchange focussed rather than examining consumption and use (Subramaniam, 2019).

This new economics of data monopolies should recognise from the outset the nature and affordances of monopoly capital; Dult (1987) argues that economics should be based instead on power relations between economic groups rather than being predicated on competition and production prices. In the new digital data age, it is not production that leads to consumption but rather the exact reverse; consumption of Datalist products and services leads to the production of ever more valuable and monetisable data for further production. A user's consumption of search, blogging platforms and image video posting platforms enables Datalists to collect and gather data that may appear to be value-less individually but is highly valuable when pulled into AI products and services. The Datalists understand that data aggregation is one of the most important monopolies they can gain in the digital economy. Those with data can reap the benefits; those without data are destined to be inputs (via their data and creative outputs) to the machines.

# 6 CONCLUSIONS

Economics must develop a theory of monopoly power based on a more detailed analysis of the technologies underlying Datalism. We have argued that we can no longer understand monopoly activities in the context of analogue assessment frameworks. However, monopoly capital theory retains relevance as an analytic and predictive tool for the real world. We propose a new research imperative for MCT, which encompasses digital technology and its impact on the relationships of economic activity, particularly a reduced focus on price and an increased focus on Datalism.

**References**
Autor D., Dorn, D., Katz, L.F., Patterson, C., Van Reenen, J., 2020, The Fall of the Labor Share and the Rise of Superstar Firms, The Quarterly Journal of

Economics, Volume 135, Issue 2, May 2020, Pages 645–709, https://doi.org/10.1093/qje/qjaa004

ADLC, (2016). Big Data, available from Joint study of the French Autorité de la concurrence and the German Bundeskartellamt on Big data, 10 May 2016: https://www.autoritedelaconcurrence.fr/en/communiques-de-presse/10-may-2016-big-data

ADLC, 2018, Sector-specific investigation into online advertising, available from: https://www.autoritedelaconcurrence.fr/en/press-release/6-march-2018-sector-specific-investigation-online-advertisin

ADLC, (2019a). Algorithms and Competition, available from: https://www.autoritedelaconcurrence.fr/sites/default/files/2019-11/2019-11-04_algorithms_and_competition.pdf

ADLC, 2019b "Digital Ecosystems, Big data and Algorithms", Issues Paper, Autoridade da Concorrência, July 2019

Auerbach, P. and Skott, P., 1988. Concentration, competition and distribution-a critique of theories of monopoly capital. *International Review of Applied Economics*, *2*(1), pp.42-61.

Baran, P. A., & Sweezy, P. M. (1966). Monopoly capital; an essay on the American economic and social order. New York, Monthly Review Press.

Baumol, W.J., 2002. Towards microeconomics of innovation: Growth engine hallmark of market economics. *Atlantic Economic Journal*, *30*(1), pp.1–12.

Battistini, A., 2013. A Theory of Profit and Competition. *Evolutionary and Institutional Economics Review*, *10*(2), pp.269-294.

Carson, L. and Richard, L., 2002. *Competition, Economic Profit, and Political Capture* (No. CEP 02-09).

Carlota Perez, Technological revolutions and techno-economic paradigms, Cambridge Journal of Economics, Volume 34, Issue 1, January 2010, Pages 185–202, https://doi.org/10.1093/cje/bep051

CCB, 2017 "Big data and Innovation: Implications for Competition Policy in Canada", Canada Competition Bureau, 18 September 2017

Coveri, A., Cozza, C. and Guarascio, D., 2021. *Monopoly capitalism in the digital era* (No. 2021/33). LEM Working Paper Series

Cowling, K. 1995. "Monopoly capitalism and stagnation." Review of Political Economy 7(4): 430–46.

Cowling and Sugden, 1998, The Essence of the Modern Corporation: Markets, Strategic Decision-Making and the Theory of the Firm, The Manchester School of Economic & Social Studies, 66, issue 1, p. 59–86.

Cowling, K. and Tomlinson, P.R., 2011. Post the 'Washington Consensus': Economic governance and industrial strategies for the twenty-first century. Cambridge Journal of Economics, 35(5), pp.831–852.

Crémer, J., de Montjoye, Y., Schweitzer, H., 2019, Competition policy for the digital era, available from: https://ec.europa.eu/competition/publications/reports/kd0419345enn.pdf

Cusumano, M.A., Gawer, A. and Yoffie, D.B., 2019. *The business of platforms: Strategy in the age of digital competition, innovation, and power*. New York: Harper Business.

Dunn, B., 2017. Against neoliberalism as a concept. *Capital & Class*, *41*(3), pp.435–454.

Dunn, B., 2017. Class, capital and the global unfree market: resituating theories of monopoly capitalism and unequal exchange. *Science & Society*, *81*(3), pp.348–374.

ECJ, 1979, Hoffman-Laroche, European Court of Justice, 13 February 1979, 85/76.

Evans, David S., Andrei Hagiu, and Richard Schmalensee. *Invisible Engines: How Software Platforms Drive Innovation and Transform Industries*. MIT Press, 2006.

Faulkner, P., Lawson, C., Runde, J., 2010, Theorising technology, Cambridge Journal of Economics, Volume 34, Issue 1, January 2010, Pages 1–16, https://doi.org/10.1093/cje/bep084

Foroohar, R., 2020, A Transatlantic Effort to Take on Big Tech, Financial Times UK Edition, 7th December 2020

Foroohar, R.,2019, *Don't be evil: How big tech betrayed its founding principles -- and all of us*, Penguin.

Foster, J.B., McChesney, R.W., Jonna, R.J., 2011, Monopoly and Competition in Twenty-First Century Capitalism, Monthly Review, April 1st 2011, Monthly Review Press

Foster and McChesney, 2012, The Endless Crisis How Monopoly-Finance Capital Produces Stagnation and Upheaval from the USA to China, May 2012, Monthly review press (New York, N.Y.: 1949) 64(1):1

Foster, J.B., Magdoff, F., 2009, The Great Financial Crisis: Causes and Consequences. NYU Press, 2009.

Franklin, U., 1989, The Real World of Technology, Volume 1989 of CBC Massey lectures, CBC Enterprises

Frenkel, S., Thompson, S., 2023, July 15, 'Not for Machines to Harvest'; Data Revolts Break Out Against A.I. *The New York Times, available from: https://www.nytimes.com/2023/07/15/technology/artificial-intelligence-models-chat-data.html,* accessed July 15, 2023

Gawer, A. and Cusumano, M.A., 2014. Industry platforms and ecosystem innovation. *Journal of product innovation management*, *31*(3), pp.417-433.

Giovanni Dosi, Marco Grazzi, On the nature of technologies: knowledge, procedures, artefacts and production inputs, Cambridge Journal of Economics, Volume 34, Issue 1, January 2010, Pages 173–184, https://doi.org/10.1093/cje/bep041

Grullon, G., Larkin, Y., Michaely, R., 2019. "Are US Industries Becoming More Concentrated?," Review of Finance, European Finance Association, vol. 23(4), pages 697-743.

IPS, 2000, Top 200: The Rise of Corporate Global Power, Institute of Public Studies, December 2000

Kharpal, A., 2017, Tech CEOs back call for basic income as AI job losses threaten industry backlash, CNBC, Feb 21 2017, available from Tech CEOs back call for basic income as AI job losses threaten industry backlash.

Lambert, T., 2019, Monopoly capital and entrepreneurship: whither small business?, Cambridge Journal of Economics 2019, 43, 1577–1595

doi:10.1093/cje/bey060

Lande, R. H., 2021, Book Review of "Antitrust: Taking on Monopoly from the Gilded Age to the Digital Age" by Senator Amy Klobuchar (June 1, 2021). The Antitrust Source, June 2021, University of Baltimore School of Law Legal Studies Research Paper Forthcoming, Available at SSRN: https://ssrn.com/abstract=3876760

Lodhi, A. (2017). 'Neoliberalism and Political Crisis': A Postulate of the Causal Dialectics Behind the Unfolding Trumpian Crisis. Critique: Journal of Socialist Theory, 45(4), 519–547.

Metcalfe, J.S., 2010, Technology and economic theory, Cambridge Journal of Economics 2010, 34, 153–171 doi:10.1093/cje/bep075

Mulligan, C.E., 2012. The Communications Industries in the Era of Convergence: Routledge Studies in Global Competition. Taylor & Francis.

Nolan, P., Zhang, J., Liu, C., 2008, The global business revolution, the cascade effect, and the challenge for firms from developing countries, Cambridge Journal of Economics, Volume 32, Issue 1, January 2008, Pages 29–47, https://doi.org/10.1093/cje/bem016

Normann, R., 2001. Reframing business: When the map changes the landscape. John Wiley & Sons.

OECD, 2016, Big data: Bringing competition policy to the digital era, 29-30 November 2016, available from: https://www.oecd.org/competition/big-data-bringing-competition-policy-to-the-digital-era.htm

OECD, 2018, Oslo Manual 2018: Guidelines for Collecting, Reporting and Using Data on Innovation, 4th Edition, ISBN 978-92-79-92581-8

Pagano, U., 2014, The crisis of intellectual monopoly capitalism, Cambridge Journal of Economics, 38, 1409–1429 doi:10.1093/cje/beu025

Pallaro, E., Subramanian, N., Abdulrahman, M.D. and Liu, C., 2015. Sustainable production and consumption in the automotive sector: integrated review framework and research directions. *Sustainable Production and Consumption*, *4*, pp.47-61.

Parker, G.G., Van Alstyne, M.W. and Choudary, S.P., 2016. Platform revolution: How networked markets are transforming the economy and how to make them work for you. WW Norton & Company.

PETIT, N., 2020 Big tech and the digital economy: the moligopoly scenario, Oxford: Oxford University Press, 2020 - https://hdl.handle.net/1814/68567

Petr, J.L., 1984. An Assault on the Citadel: Is a constructive synthesis feasible? *Journal of Economic Issues*, *18*(2), pp.589–597.

Pisoni, A., Onetti, A., 2018, When startups exit: comparing strategies in Europe and the USA, Journal of Business Strategy

Pitelis, C.N., 2016. Learning, innovation, increasing returns and resource creation: Luigi Pasinetti's 'original sin and call for post-classical economics. *Cambridge Journal of Economics*, *40*(6), pp.1761-1786.

PITELIS, C. N. & TOMLINSON, P. R. 2017. Industrial organisation, the degree of monopoly and macroeconomic performance – A perspective on the contribution of Keith Cowling (1936–2016). *International Journal of Industrial Organization,* 55**,** 182–189.

Robinson J., 1933, "The Economics of Imperfect Competition", London, MacMillan, 1933

Roose, K., 2021, Futureproof: 9 Rules for Humans in the Age of Automation, John Murray Press

Rosenberg, Nathan, Inside the Black Box: Technology and Economics (1982). University of Illinois at Urbana-Champaign's Academy for Entrepreneurial Leadership Historical Research Reference in Entrepreneurship, Available at SSRN: https://ssrn.com/abstract=1496197

Rikap, C. and Lundvall, B.Å., 2020. Big tech, knowledge predation and the implications for development. *Innovation and Development*, pp.1-28.

Rikap, C., 2021. *Capitalism, power and innovation: Intellectual monopoly capitalism uncovered*. Routledge.

Sanchez-Cartas, J. M., 2020 "The Panzar–Rosse H Statistic and Monopoly. Issues on its Use as a Market Power Measure" *The B.E. Journal of Economic Analysis & Policy*, vol. 20, no. 4, 2020, pp. 20200193. https://doi.org/10.1515/bejeap-2020-0193

Sawyer, M., 1988, Theories of Monopoly Capital, Journal of Economic Surveys, 1988, vol. 2, issue 1, 47-76

Smith, T., 2010, Technological change in Capitalism: some Marxian themes, Cambridge Journal of Economics, Volume 34, Issue 1, January 2010, Pages 203–212, https://doi.org/10.1093/cje/bep048

Konstantina Spanaki, Zeynep Gürgüç, Richard Adams & Catherine Mulligan (2018) Data supply chain (DSC): research synthesis and future directions, International Journal of Production Research, 56:13, 4447-4466, DOI: 10.1080/00207543.2017.1399222

Subramaniam, M. (2020). Digital ecosystems and their implications for competitive strategy. *Journal of Organization Design, 9*(1), 1-10. doi:10.1186/s41469-020-00073-0

Tsoulfidis, L., 2010. *Competing schools of economic thought*. Springer Science & Business Media.

UK Government, 2019, Unlocking digital competition Report of the Digital Competition Expert Panel, ISBN 978-1-912809-44-8

US House of Representatives, 2020, Investigation of Competition in Digital Markets, available from: https://judiciary.house.gov/uploadedfiles/competition_in_digital_markets.pdf

Wang, N., Heng, X., & Grossklags. J., 2011, Third-party apps on Facebook: privacy and the illusion of control. In *Proceedings of the 5th ACM Symposium on Computer-Human Interaction for Management of Information Technology* (*CHIMIT '11*). Association for Computing Machinery, New York, NY, USA, Article 4, 1–10 DOI: https://doi.org/10.1145/2076444.2076448

Waters, M., 1995. The thesis of the loss of the perfect market. British Journal of Sociology, pp.409–428.

Waters, R., 2017, Silicon Valley aims to engineer a universal basic income solution: UBI conjures up an image of an ideal future that appeals to techies, Financial Times, May 3rd 2017, available from: https://www.ft.com/content/b0659404-0fea-11e7-a88c-50ba212dce4d

World Bank, 2010, The World Bank and Microsoft Sign Partnership Agreement to Promote Development in Africa, available at: https://www.worldbank.org/en/news/press-release/2010/01/31/the-world-bank-and-microsoft-sign-partnership-agreement-to-promote-development-in-africa

FTC, 2021, Monopolization Defined, US Federal Trade Commission, available from: https://www.ftc.gov/tips-advice/competition-guidance/guide-antitrust-laws/single-firm-conduct/monopolization-defined

USSC, 1956, US Supreme Court, United States v. E. I. du Pont de Nemours & Co., 351 U.S. 377 (1956)